# Flow Regime Transition in Countercurrent Packed Column Monitored by ECT

Zhigang Li, Yuan Chen, Yunjie Yang, Chang Liu, Mathieu Lucquiaud and Jiabin Jia[*]

School of Engineering, The University of Edinburgh, Edinburgh, UK, EH9 3JL

jiabin.jia@ed.ac.uk

**Abstract**

Vertical packed columns are widely used in absorption, stripping and distillation processes. Flooding will occur in the vertical packed columns as a result of excessive liquid accumulation, which reduces mass transfer efficiency and causes a large pressure drop. Pressure drop measurements are typically used as the hydrodynamic parameter for predicting flooding. They are, however, only indicative of the occurrence of transition of the flow regime across the packed column. They offer limited spatial information to mass transfer packed column operators and designers. In this work, a new method using Electrical Capacitance Tomography (ECT) is implemented for the first time so that real-time flow regime monitoring at different vertical positions is achieved in a countercurrent packed bed column using ECT. Two normalisation methods are implemented to monitor the transition from pre-loading to flooding in a column of 200 mm diameter, 1200 mm height filled with plastic structured packing. Liquid distribution in the column can be qualitatively visualised via reconstructed ECT images. A flooding index is implemented to quantitatively indicate the progression of local flooding. In experiments, the degree of local flooding is quantified at various gas flow rates and locations of ECT sensor. ECT images were compared with pressure drop and visual observation. The experimental results demonstrate that ECT is capable of monitoring liquid distribution, identifying flow regime transitions and predicting local flooding.

**1. Introduction**

Countercurrent flow packed column plays a fundamental role in many chemical unit operations, such as $CO_2$ absorption [1], where a gravity-driven falling liquid film flows downwards along the packing walls in the presence of an upward flowing gas at a constant pressure gradient. Flooding poses a high risk for countercurrent flow in a vertical packed column, as it not only seriously affects the efficiency of mass transfer, but also brings other adverse consequences, such as pushing liquid into the gas pipes and then damaging the equipment. A higher pressure occurring at the bottom of the column and connected pipes can result in equipment damage, pipe leakage and even explosion. Features used to identify flooding regime include the appearance of flow reversal, excessive entrainment, a sharp rise in pressure drop, a sharp rise in the liquid hold-up, and a sharp fall in efficiency [2]. Accurate prediction of the onset of flooding is of critical importance in this regard. Therefore, it is a common practice to design and operate absorption columns below the flooding point, and get the best mass transfer with lower gas velocities and thus lower pressure drops. Lower gas velocities effectively result in the oversizing of packed column diameters, and associated increase in capital costs. With the development of new, real-time, diagnosis techniques of flooding, possibly be implemented directly at the process scale, it may be possible to gain operational insights to operate packed columns closer to the flooding point safely. With the development of real-time images of liquid distribution across vertical sections of packing, it may also be possible to observe and correct suboptimal liquid distribution.

Prevention of flooding in structured packing has been the subject of many experimental and numerical studies. Some studies optimised packing sizes and operating parameters [3] or packing arrangement [4] to push the flooding limits towards much higher throughputs in countercurrent mode. However, these practices have come at the expense of reducing the mass transfer rate. Furthermore, flooding still occurs accidentally even if the above measures are taken. Therefore, a safety factor, typically a percentage of the flooding velocity, is implemented to select operating gas velocities when sizing packed columns and to avoid detrimental effects on absorption efficacy. In order to prevent the operation of packed columns from flooding, various empirical and theoretical flooding models have been used to predict the gas velocity at the flooding point [5, 6]. However, the prediction accuracy always depends on empirical parameters related to the packed column under consideration, which is difficult to obtain. The inability to accurately predict and prevent flooding may result in loss of operating hours, decrease of product purity, equipment damages, and safety hazards, and, once large safety factors are implemented, the oversizing of column diameters.

Pressure drop, liquid hold-up and packed flooding are closely linked [2], while it is generally recognized that the pressure drop inside the column is the most crucial variable for flooding prognosis [7, 8]. However, it is difficult to localize flooding based on pressure drop. Other studies have investigated the flooding prognosis methods in order to monitor flooding set up on a packed column, e.g. gamma radiation [9] and acoustic method [10]. The acoustic method can diagnose flooding, but cannot provide information on the bypassing of liquid or the maldistribution of fluid in the column. Maldistribution of liquid flow is the most detrimental in terms of separation efficiency [11-13]. The gamma radiation method can acquire the image of the distribution of liquid and gas cross the section precisely. However, it suffers from high operation costs and radiation hazards. It is therefore used in very specific settings where measurements take place within a confined enclosure capable of stopping radiations.

Commonly applied in multiphase flow measurement, process tomography has evolved significantly over the past decades. Process tomography has many advantages, such as non-radiation, non-invasiveness and low operation cost. ECT was employed previously by Wongkia et al [14] in order to study flooding capacity of countercurrent gas-liquid flow in inclined packed beds equipped with small packings. Similarly, ECT was used to investigate pulse flow and pulse velocity in co-current trickle bed reactors [15], solid-phase distributions in a gas-solid fluidized bed [16, 17] and determination of the onset of bubbling and slugging in a fluidized bed [18]. Although ECT has been successfully used to measure liquid hold-up and distribution inside the packed column [19-21], little attention has been paid so far to the imaging of liquid distribution under flooding phenomenon, which is the motivation of this study. This study had two purposes: (1) a qualitative study of ECT images to discover liquid distribution in a countercurrent flow packed column at different flow regime; (2) a quantitative study of capacitance data provided by the ECT to evaluate a relationship between flooding and a flooding index ($P_f$). Pressure drop is monitored as a secondary measurement technique to confirm the validity of ECT method. In this paper, an ECT based process monitoring technique is proposed to study the transition process of flow regimes in the reaction column packed by structured packing. The evolution from pre-loading to flooding status in the cross-sectional liquid distribution along the axial direction of the column is captured and analysed. To the best of our knowledge, it is the first time that a flooding index is used to predict flooding locally across vertical sections of packing.

## 2. Principle of monitoring flooding with ECT

*2.1 Principle of ECT*

ECT was employed to measure the liquid distribution of the liquid hold-up inside the packed column. The methodology used in this work has been fully characterized by Wu et al. [21], which investigated the intrinsic performances of the apparatus and accuracy of the reconstruction algorithm. The ECT system [22] shown in Fig. 1, consists of sensors, control circuits and a computer with imaging software. The electrodes are made of copper foil. Eight electrodes of 10 cm length are mounted outside packed column and enclosed by an earthed guard electrode. The excitation signal is a sine wave voltage with 14 Vp-p and 200 kHz frequency. In between the electrodes and the computer is a data acquisition system which is a control circuit that allows all acquired signals are conditioned by the C/V circuit [23] and transmitted to the computer through USB 2.0. The frame rate of the ECT system is 714 frames per second. The maximum signal-noise-ratio (SNR) is 76.73 dB, and the minimum SNR is 62.25 dB. The ECT is connected to a computer where visualization of the real-time images can be accessed.

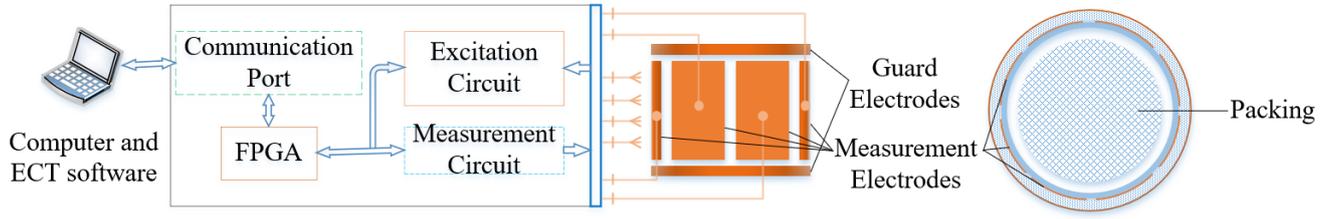

Fig. 1. ECT system and the connection to electrodes

ECT measures capacitances between all electrode pairs and then the data together with a pre-calculated sensitivity map are employed to estimate the cross-sectional permittivity change within the sensing area. The linearized model describing the relationship between normalized capacitance $C_{norm}$ and the normalized permittivity change $g$ is expressed as:

$$C_{norm} = Sg \qquad (1)$$

where $S$ is the sensitivity matrix, giving a sensitivity map for each electrode pair.

The measured capacitance $C_{mea}$ was normalized to $C_{norm}$ between a low and a high permittivity limit, i.e.

$$C_{norm} = \frac{C_{mea} - C_L}{C_F - C_L} \qquad (2)$$

where $C_L$ and $C_F$ are the reference capacitance when the region of interest is filled with low permittivity media (air) and high permittivity media (water) respectively. The low permittivity limit is set with a packed column empty of gas and liquid. The high permittivity limit is set with the packed column filled with liquid, and the cross-section of packing being submerged. Water has a relative permittivity of approximately 80. In contrast, the relative permittivities of plastic packing material and air are approximately 2.2 and 1, respectively.

*2.2 Experimental setup and methodology*

*2.2.1 Experimental setup*

A vertical packed column was built to monitor countercurrent gas-liquid two-phase flow with ECT sensor. The schematic diagram of the experimental rig is shown in Fig. 2. The external diameter of the polypropylene column is 200 mm and the height of the column is 1200 mm. The thickness of the pipe is 5 mm. The column is filled with four sections of plastic structured packings Mellapak 250.Y manufactured by Sulzer Chemtech Ltd (see Fig. 3). The diameter of packing is 180 mm with the heights of packing as 315.0 mm, 157.5 mm, 157.5 mm and 315.0 mm from top to bottom, respectively. The void fraction of polypropylene Sulzer Mellapak 250 Y is 12% based on the manufacturer specifications. All four structured packings have the same porosity. The packing geometry is characterized by an inclination angle of the flow channels with a horizontal direction of 45°. The packing elements, made of plastic, were alternatively rotated around the axis of the column by 90° relative to each other in order to encourage uniform liquid distribution.

The column was operated in a countercurrent flow configuration with liquid entering at the top of the column using a liquid sprayer and the air entering at the bottom using a gas distributor. The liquid sprayer used for liquid feeding has a diameter of 9 cm, and orifices of 4 mm in diameter. Liquid flow is controlled with an adjustable bypass valve with the liquid loads ranging from 13.4 to 38.9 $m^3/(m^2h)$ and monitored with an electromagnetic flowmeter (OMEGA, FMG71B-A-BSP, with an accuracy of ± 2.0%). Two blowers (Windjammer, 119153) introduce air into the column from the bottom. The air volumetric flow rate was increased from 0.0104 to 0.0226 $m^3/s$ in small increments.

Liquid distributions were imaged via the ECT sensor at various longitudinal positions illustrated in the orange zones of Fig. 2 (Bottom = 0 mm, Lower = 155 mm, Middle = 325 mm and Upper = 495 mm). The locations of the measurements were chosen to examine the liquid distribution along with the column height. Two pressure transducers were placed at the top and bottom of the column to measure the pressures drop, as shown in Fig. 2.

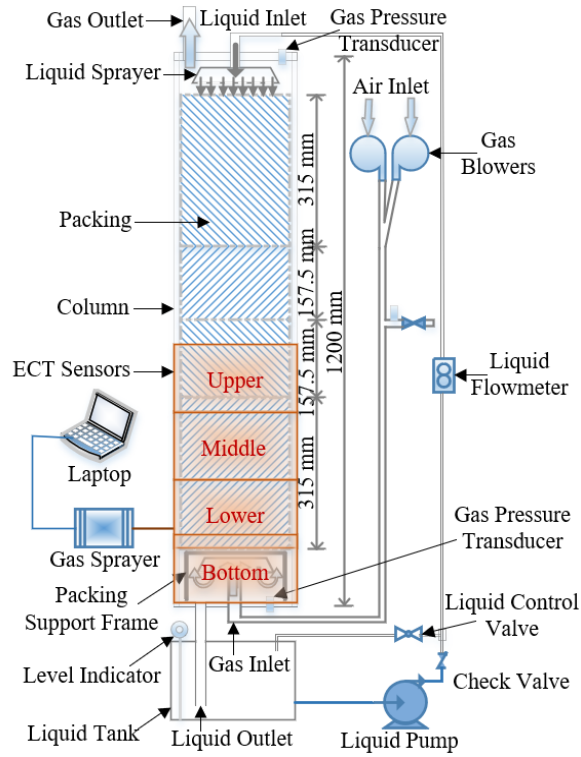

Fig. 2. Schematic diagram of the experimental rig

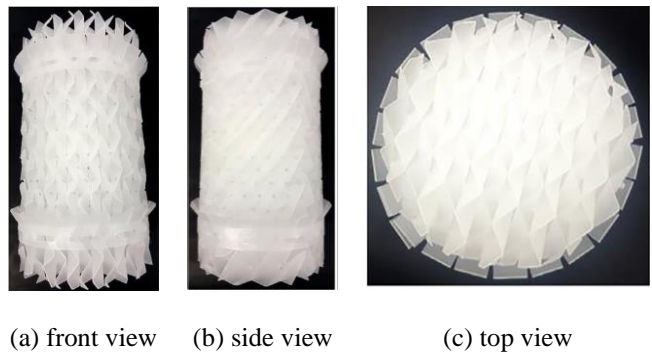

(a) front view    (b) side view    (c) top view

Fig. 3. Mellapak 250Y packing. Reprinted from [21]. Copyright © (2018) Elsevier. Reprinted with permission.

*2.2.2 Experimental procedures*

The reference capacitances ($C_l$ and $C_F$) were measured first. As mentioned above, $C_l$ is the capacitance when the column is filled with plastic structured packing and stagnant air, as shown in Fig. 4(a). $C_F$ is the capacitance when the packed column is filled with water, as shown in Fig. 4(b). Experiment results in section 3.2 are obtained with this normalisation method. The liquid flow rate was set at a constant value, the gas flow rate was adjusted to the desired value. The real status of the column was photographed, and ECT images of the column were reconstructed. The gas flow rates were increased in the smallest increments of air volumetric flow rate until the column is flooded. At each gas flow rate, after calibration measurement for ECT was taken, ECT data within 20 seconds at each gas flow rate were recorded and averaged for offline data analysis. Meanwhile, pressures at the top and bottom of the column are measured simultaneously.

Procedure in making simultaneous measurements of pressure drop and ECT images was similar except that the ECT at each flow condition were the results of an independent run using two different calibration methods. In another normalisation method, as shown in Fig. 4(c), capacitance in flooding condition $C_f$ is used as high permittivity limit to replace $C_F$ in Eq. (2). Experiment results in subsection 3.3 are obtained with this normalisation method. To avoid inaccuracy of capacitance reference, $C_f$ is the average of 20s measurement, rather than a single measurement, so that the variation of the normalised capacitance $C_{norm}$ in Eq.(2) could be restricted.

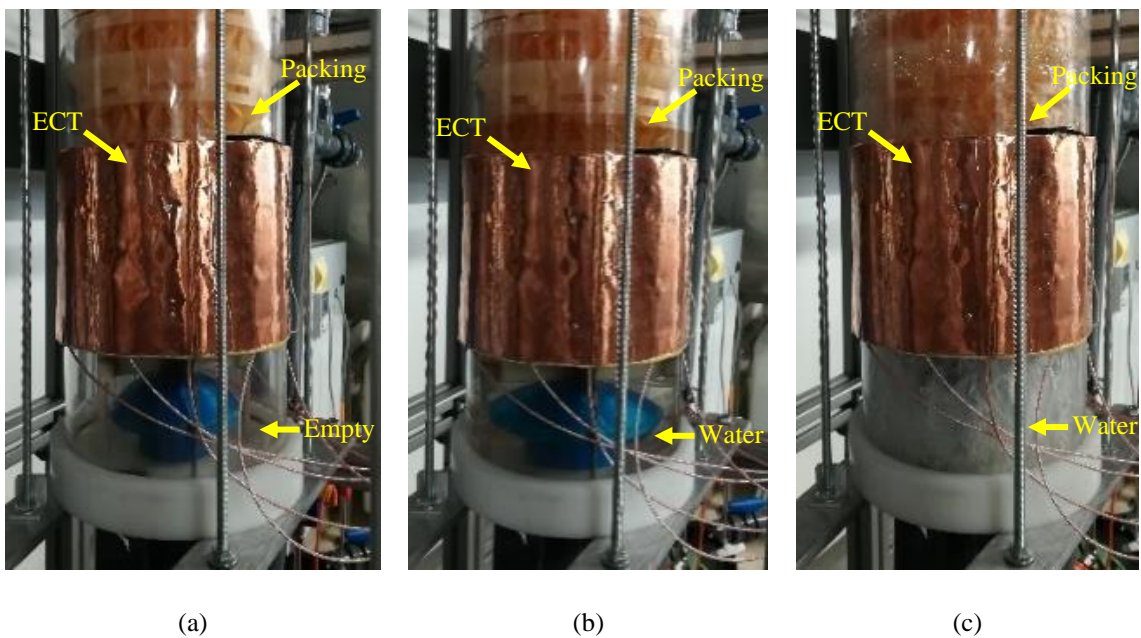

(a)          (b)          (c)

Fig. 4. Photos of the flow within a structred packing for different operting condtions

(a) $C_{l(j)}$ empty column condition, (b) $C_{F(j)}$ full water condition and (c) $C_{f(j)}$ flooding condition

*2.3 Flooding index ($P_f$) calculation model*

Liquid distribution, liquid hold-up and packing efficiency are significantly affected by different operation regime in the structured packing column. There are three main operation regimes, i.e. pre-loading regime, loading regime and flooding regime [24]. In the pre-loading regime, liquid flows freely into the structured packing and form a liquid film on the structured packing surface. The loading point is a critical point where the vapour kinetic energy is sufficient to destabilize the liquid film and influence the fluid flow. When the operating column transitions from pre-loading to the loading regime, both liquid and gas flow rate starts to affect the liquid film formed over the packing [24]. A further increase of liquid or gas flow rate will then result in fluid accumulation at the bottom of the bed, which indicates the column has reached loading regime. When liquid accumulation approaches the top of the column, the column has reached a state of flooding.

$P_f$ is used here to refer to the degree of local flooding or the local maximum liquid capacity of the packed column. The liquid entrainment may become excessive if the $P_f$ at a certain place is too high. The real-time liquid distributions were imaged at various longitudinal positions using the Linear Back Projection (LBP) algorithm [25] based on $C_{norm}$ in Eq. (2). To evaluate the stage of flooding quantitatively, the $P_f$ is formulated in Eq. (3), where $V(j)$ is the value of normalized permittivity at the $j$th pixel. $M$ is the total number of the pixels of a reconstructed image. $M$ is 3228 in this work. $P_f$ represents the degree of a state approaching local flooding.

$$P_f = \frac{\sum_{j=1}^{M} V(j)}{M} \qquad (3)$$

In addition to pressure drop, $P_f$ and ECT images of liquid distribution are able to quantitatively and qualitatively indicate the developing process of flooding in the packed column.

**3. Results and discussions**

*3.1 Pressure drop*

The operating regions of a packed column can be identified, in principle, via visual inspection of a transparent column. The majority of industrial columns is, however, non-transparent. Monitoring pressure drop along the column is the conventional way of monitoring loading and flooding. Pressure drop in the packed column can be estimated using the empirical models [26, 27]. In the experiments, the liquid rate remained at 38.9 m$^3$/(m$^2$ h). Pressure drop is plotted in Fig. 5 against the increase of air volumetric flow rate from 0.0104 m$^3$/s to 0.0219 m$^3$/s. As illustrated in Fig. 5, it is difficult to directly identify the loading point on the discrete pressure drop points and the regression curve, $f(x) = 2.346 \cdot 10^5 \cdot x^3 - 1.015 \cdot 10^4 \cdot x^2 + 151.1 \cdot x - 0.6468$. The gradient of the pressure drop curve is $f(x) = 7.2 \cdot 10^{10} \cdot x^4 - 4.375 \cdot 10^9 \cdot x^3 + 9.828 \cdot 10^7 \cdot x^2 - 9.661 \cdot 10^5 \cdot x + 3511$. The gradient rapidly increases from 0.0196 m$^3$/s, which indicates the loading point of the packed column. After the loading point is reached, liquid starts to accumulate at the bottom of the column. The operation of the packed column changes from the loading regime to flooding regime. The continuous increase of gas flow rates causes higher liquid hold-up, which in turn causes a higher pressure drop and liquid starts bypassing the packing bed [4]. As shown in the photos in Table 1 in subsection 3.2, the local accumulation of liquid can be clearly observed at the bottom of the column, and the column is slowly "drowned" in the liquid.

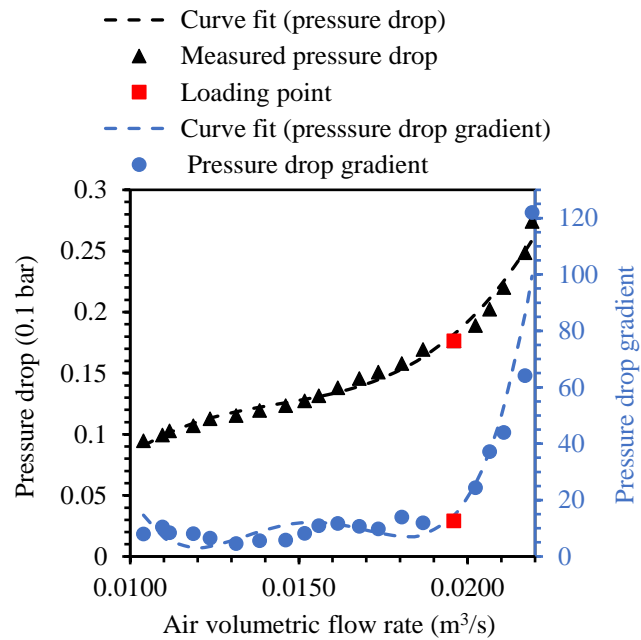

Fig. 5. Dependence of pressure drop and gradient of pressure drop on air volumetric flow rate

*3.2 Transition of flow regime*

The flow regimes in the countercurrent packed column are photographed, while ECT images are reconstructed over the column cross-section for the bottom position without packing and the lower position with structured packing. The ECT sensor was removed for visual observation in the structured packing area. In this series of experiments, high capacitance reference was taken from the column filled with water, in order to monitor the transition process of gas-liquid distribution from pre-loading regime to flooding regime. Table 1 presents the measured liquid distribution at various air volumetric flow rate from 0.0138 m$^3$/s to 0.0217 m$^3$/s. Red areas in the images represent higher liquid hold-up, while the blue areas represent lower liquid hold-up. ECT images are analyzed separately in terms of the bottom and lower section of the column.

Initially, a smooth falling liquid film could be seen when the air volumetric flow rate was lower than 0.0162 m$^3$/s. Gradually, the liquid film with bubbles became turbulent. Hydrodynamic instability corresponds to a transition from pre-loading regime to the loading regime, which associates with a significant increase of liquid hold-up [28]. In the pre-loading regime with air volumetric flow rate from 0.0138 m$^3$/s to 0.0162 m$^3$/s, The ECT images of the bottom section show that the liquid hold-up and distribution patterns are stable. The liquid is uniformly distributed around the inner area of the column, and liquid hold-up at the inner area is slightly higher than that of the peripheral area. With the increase of air volumetric flow rate from 0.0162 m$^3$/s onwards, visual observations during the experiments also confirm that there are droplets entrainment and flow reversal occurs in the packed column. Red patches then red ring appear on the boundaries of ECT images, which means liquid is accumulated with a larger accumulation near the wall of the column than that in the centre of the column.

Structured packing in the lower section prolongs residence time of the liquid phase. Therefore, at any air volumetric flow rate up to 0.0162 m$^3$/s, the liquid hold-up in the lower section is higher than that in the bottom section. After the loading point is passed, the liquid flow across the lower section remains evenly distributed, namely, more liquid is homogeneously distributed in the packing area. Less liquid is distributed around the gap between packing and the column wall than in the middle of the section. It implies that gas-liquid mass transfer will take place in the centre of the packing rather than the void space between the packing and column wall. Compared with the bottom section, the flooding has a more dramatic formation process in the lower section. A sudden transfer from the loading regime to flooding regime takes place from 0.0206 m$^3$/s to 0.0211 m$^3$/s. More and more liquid accumulates near the wall of the column. Since the liquid bypasses structured packing, the liquid hold-up of the peripheral area is higher at 0.0217 m$^3$/s. By now, the gas kinetic energy is sufficient to destabilize the liquid film and reverse the liquid flow, which leads to column flooding. These results demonstrate the advantage of using ECT over a conventional pressure drop method, as ECT can provide detailed spatial information at different stages of the flow regime in the packed column.

Table 1 Air volumetric flow rate, visual observation of the column and typical reconstructed ECT images.

| Air volumetric flow rate (m$^3$/s) | 0.0138 | 0.0146 | 0.0152 | 0.0156 | 0.0162 | 0.0168 | 0.0173 |
|---|---|---|---|---|---|---|---|
| Countercurrent flow | 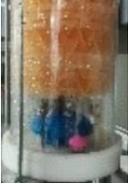 | 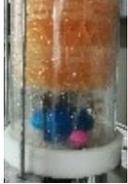 | 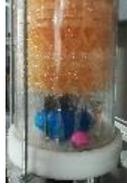 | 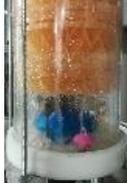 | 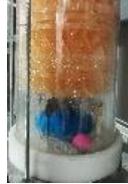 | 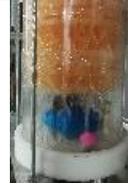 | 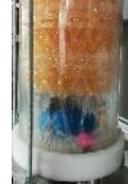 |
| Reconstructed Images (Lower) | 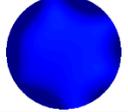 | 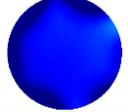 | 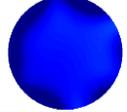 | 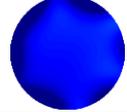 | 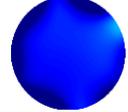 | 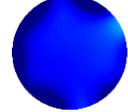 | 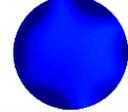 |
| Reconstructed Images (Bottom) | 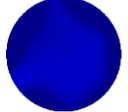 | 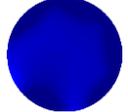 | 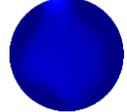 | 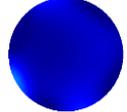 | 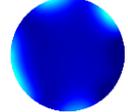 | 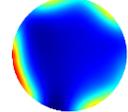 | 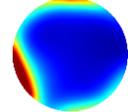 |
| **Air volumetric flow rate (m$^3$/s)** | **0.0180** | **0.0187** | **0.0196** | **0.0202** | **0.0206** | **0.0211** | **0.0217** |
| Countercurrent flow | 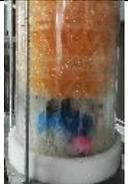 | 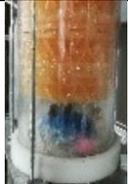 | 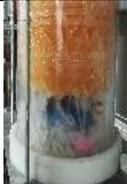 | 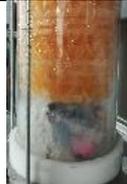 | 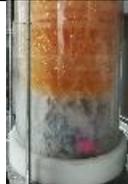 | 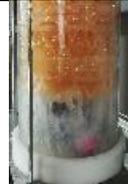 | 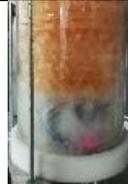 |
| Reconstructed Images (Lower) | 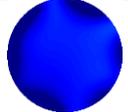 | 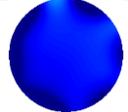 | 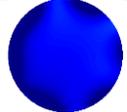 | 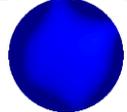 | 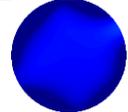 | 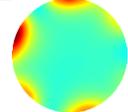 | 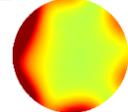 |
| Reconstructed Images (Bottom) | 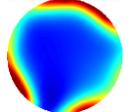 | 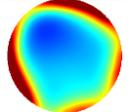 | 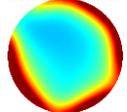 | 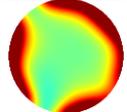 | 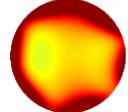 | 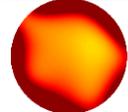 | 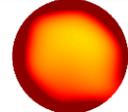 |
| Colour map | 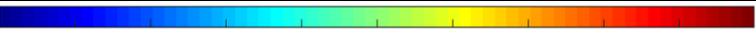 | | | | | | |

*3.3 Quantification of local flooding*

Monitoring flooding with ECT allows for a spatial determination of liquid distribution compared to a reference state. We introduce the flooding index $P_f$ determined in Eq. (3) to represent the degree of local flooding. $P_f =1$ implies that the cross-section of packing has reached a state of accumulation of liquid corresponding to flooding. In this series of experiments, the high capacitance reference was taken when the column was flooded. Reconstructed images and the corresponding $P_f$ are shown in Table. 2. Reconstructed images illustrate the liquid distribution at different column height against air volumetric flow rate. As air volumetric flow rate increases, ECT images at the lower, middle and upper locations display the same transition before and after local flooding. For each section of packing, the structured packing holds liquid preferentially towards the centre of the column, before flooding occurs. When the air volumetric flow rate reaches certain values, respectively 0.0224 m$^3$/s, 0.0219 m$^3$/s, and 0.0211 m$^3$/s for the upper, middle and lower sections, the column is flooded quickly. As expected, local flooding starts first in the lower section of packing, then propagates to the middle and upper locations. Under this set of experiments, it is possible to extrapolate that the ECT sensor should be installed at the bottom position if the loading point needs to be identified early. If the flooding point needs to be identified, ECT sensor should be installed at the top position of the column. Though tomographic images have relatively low resolution, they are still able to indicate liquid distribution in the packed column, which we believe is valuable information for the opaque reaction columns.

**Table 2** Typical reconstructed images corresponding to different sections

| | Air volumetric flow rate (m³/s) | 0.0187 | 0.0196 | 0.0202 | 0.0206 | 0.0211 | 0.0217 | 0.0219 | 0.0224 | 0.0226 |
|---|---|---|---|---|---|---|---|---|---|---|
| Upper | Reconstructed images | | | | | | | | | |
| | Flooding index $P_f$ | 0.0278 | 0.0269 | 0.0249 | 0.0196 | 0.0120 | 0.0020 | 0.0474 | 0.7318 | 1 |
| Middle | Reconstructed images | | | | | | | | | |
| | Flooding index $P_f$ | 0.0397 | 0.0369 | 0.0295 | 0.0211 | 0.0120 | 0.3855 | 0.8911 | 1 | 1 |
| Lower | Reconstructed images | | | | | | | | | |
| | Flooding index $P_f$ | 0.1626 | 0.1316 | 0.0814 | 0.1104 | 0.8095 | 1 | 1 | 1 | 1 |
| Colour map | | 0    0.1    0.2    0.3    0.4    0.5    0.6    0.7    0.8    0.9    1 | | | | | | | | | |

The calculated $P_f$ values were compared with a pressure drop measured. Fig. 6 shows the influence of gas velocity on the $P_f$ at the bottom of the column and pressure drop gradient. It can be observed that the $P_f$ at the column bottom starts to increase from 0.0152 m³/s air volumetric flow rate, while the pressure drop gradient has an obvious increase after 0.0196 m³/s air volumetric flow rate. It is worth noting that the pressure drop reports the change of pressure across the column, but that the ECT monitors the flooding index locally in the bottom section of the column. These results indicate that $P_f$ obtained using ECT can provide more sensitive and accurate information for an early prognosis of flooding. It is also possible to detect flooding locally, and, by extension, increase confidence in operating with gas velocities closer to the column flooding velocity.

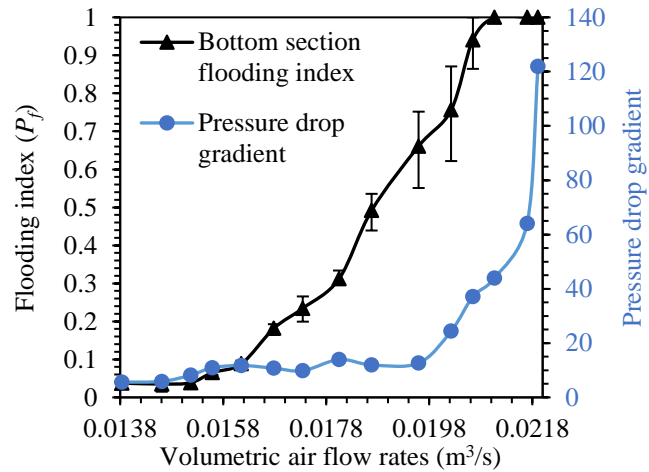

Fig. 6. $P_f$ at the bottom section and pressure drop gradient

Fig. 7 illustrates the influence of the air volumetric flow rate to the Pf at lower, middle and upper sections of the column. Three Pf curves of lower, middle and upper position have a similar developing trend. As the column is sequentially flooded across the lower section, the middle position and the upper position, three Pf curves reach 1 sequentially. During this flooding process, the pressure drop against the air volumetric flow rate is plotted in Fig. 7 for clarity. There is no apparent inflection point on the pressure drop curve. In contrast, $P_f$ extracted from ECT image can quantitatively indicate a degree of local flooding and reveal a sudden appearance of local flooding, for example, $P_f$ 0.11 at 0.0206 m$^3$/s air volumetric flow rate in the lower position. The transition of flooding took place rapidly and intensively between 0.0206 m$^3$/s to 0.0217 m$^3$/s air volumetric flow rate. Large $P_f$ variation at 0.0211 m$^3$/s reflects severe flow turbulence, which can be visually observed on the photos in Table 1. The same phenomenon is demonstrated on the curve of the middle and upper section.

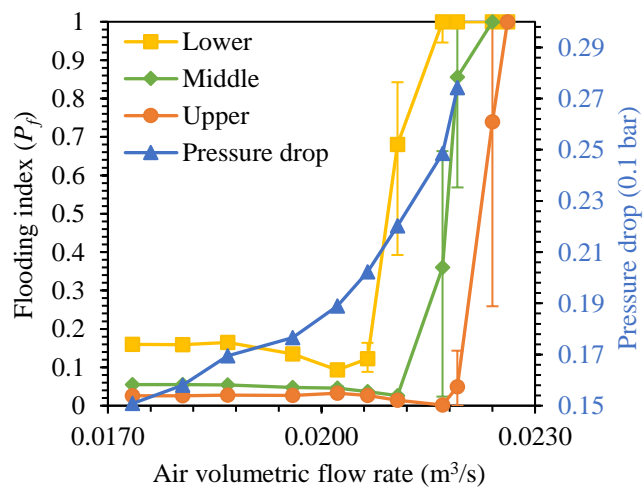

Fig. 7. $P_f$ of the lower, middle, upper sections and pressure drop

The standard deviation of flooding index in Fig 6 and Fig 7 was caused by turbulent flow itself. High system speed of ECT (714 frames per second) enables a fast snapshot of one individual cross-sectional image to truly reflect the nature of turbulent flow in the flooding process. It is worth mentioning another phenomenon. The $P_f$ always goes down first before flooding occurs, which indicates that the liquid hold-up decreases then increases suddenly towards flooding. Take the lower section as an example again, the $P_f$ drops from 0.16 to 0.09, then rises to 1.0. This phenomenon is consistent with results obtained in previous studies [29]. This phenomenon was repeatedly captured at all three sections. These results suggest that the use of $P_f$ obtained using ECT can provide new insights into column operation for early prognosis of local flooding compared to traditional pressure drop method.

## 4. Conclusion

Flooding monitoring in packed columns is demonstrated as a novel application of Electrical Capacitance Tomography (ECT). ECT was used to investigate liquid distribution across three sections of packing in a countercurrent gas-liquid column equipped with plastic Mellapak 250.Y structured packing under different operating conditions of gas velocity. Reconstructed images by ECT can provide a spatial image with the distribution of liquid across different sections of a column, thus indicating the spatial formation process of flooding. Because ECT is a non-intrusive, portable imaging method, it could be deployed at the process scale to optimise column performance. The flooding index inferred from real-time measurements is a quantitative method to evaluate the degree of local flooding. The experimental results demonstrate that flooding can be predicted by using the flooding index curve. For increasing gas velocities, the onset of flooding is preceded by a drop of the value of the flooding index. The new quantitative method based on ECT imaging provides additional insights for flooding monitoring in addition to monitoring pressure drop. This study opens up a new method of flooding prognosis with ECT in the packed reaction columns used in chemical engineering applications.

**Acknowledgement**

This is a follow-up work after the project EP/M001482/1 supported by the United Kingdom Engineering and Physical Sciences Research Council (EPSRC). The authors would like to express their gratitude for the comments and suggestion offered by the reviewers.